\documentclass[aps,prl,reprint,10pt,superscriptaddress]{revtex4-1}

\usepackage{color}
\usepackage{bbm}
\usepackage{amsmath,amsfonts,amsthm}
\usepackage{graphicx}
\usepackage[colorlinks,linkcolor=black,urlcolor=black,citecolor=black]{hyperref}

\newtheorem{theorem}{Theorem}
\newtheorem{lemma}[theorem]{Lemma}

\newcommand{\EE}{\mathbb{E}}
\newcommand{\CC}{\mathbb{C}}
\newcommand{\calU}{\mathcal{U}}
\newcommand{\calE}{\mathcal{E}}
\newcommand{\calL}{\mathcal{L}}

\newcommand{\ket}[1]{|{#1}\rangle}
\newcommand{\bra}[1]{\langle{#1}|}
\newcommand{\braket}[2]{\langle{#1}|{#2}\rangle}

\newcommand{\Ket}[1]{\bigl|{#1}\bigr)}
\newcommand{\Bra}[1]{\bigl({#1}\bigr|}
\newcommand{\Braket}[2]{\bigl({#1}\bigr|{#2}\bigr)}

\providecommand{\abs}[1]{\lvert#1\rvert} 
\providecommand{\norm}[1]{\lVert#1\rVert}

\newcommand{\set}[1]{\lbrace #1 \rbrace}

\newcommand{\eref}[1]{Eq.~\eqref{#1}}
\renewcommand{\vec}[1]{\mathbf{#1}}

\DeclareMathOperator{\Tr}{tr}
\DeclareMathOperator{\Var}{Var}

\DeclareMathOperator{\poly}{poly}

\begin{document}

\title{Direct Fidelity Estimation from Few Pauli Measurements}

\author{Steven T.\ Flammia}\affiliation{Institute for Quantum Information, California Institute of Technology}

\author{Yi-Kai Liu}\affiliation{Computer Science Department, University of California, Berkeley}

\date{April 29, 2011}

\begin{abstract}
We describe a simple method for certifying that an experimental device prepares a desired quantum state $\rho$.  Our method is applicable to any pure state $\rho$, and it provides an estimate of the fidelity between $\rho$ and the actual (arbitrary) state in the lab, up to a constant additive error. The method requires measuring only a \emph{constant} number of Pauli expectation values, selected at random according to an importance-weighting rule. Our method is faster than full tomography by a factor of $d$, the dimension of the state space, and extends easily and naturally to quantum channels. 
\end{abstract}

\maketitle

In recent years there has been substantial progress in preparing many-body entangled quantum states in the laboratory \cite{quantum-coherence}.  
A key step in such experiments is to \emph{verify} that the state of the system is the desired one.  This can be done using quantum state tomography, or techniques such as entanglement witnesses \cite{Guhne2009}.  However, in many cases these solutions are not fully satisfactory.  Tomography gives complete information about the state, but it is very resource-intensive, and has difficulty scaling to large systems.  Entanglement witnesses can be much easier to implement, but are not a generic solution since known constructions only work for special quantum states.

Here we propose a new method, \textit{direct fidelity estimation}, that is much faster than tomography, is applicable to a large class of quantum states, and requires minimal experimental resources.  Let us first describe the setting of the problem.  Consider a system of $n$ qubits, with Hilbert space dimension $d = 2^n$, and let $\rho$ be the desired state, i.e.\ the state we hope to accurately prepare. We make two basic assumptions.  First, we assume that $\rho$ is \emph{pure}.  However, we do not assume any additional structure or symmetry, so our method goes beyond previous work \cite{Somma06, Guhne07} to encompass nearly all of the states of interest in experimental quantum information science (e.g., the GHZ and W states, stabilizer states, cluster states, matrix product states, projected entangled pair states, etc.) in a unified framework.  Second, we assume that we can measure $n$-qubit Pauli observables, that is, tensor products of single-qubit Pauli operators; we do not need to perform any other operations.  Thus our method is applicable to any system that is capable of single-qubit gates and readout, without needing to rely on 2-qubit gates or entangled measurements.

Our method works by measuring a random subset of Pauli observables chosen according to an ``importance-weighting'' rule.  Roughly, we select Pauli operators that are most likely to detect deviations from the desired state $\rho$.  We use the resulting measurement statistics to estimate the fidelity $F(\rho,\sigma)$, where $\sigma$ is the actual state in the lab. Surprisingly, although there are $4^n$ distinct Pauli operators, we only need to sample a \emph{constant} number of them to estimate $F(\rho,\sigma)$ up to a constant additive error, for \emph{arbitrary} $\sigma$. That is, for every possible state $\sigma$, with high probability over the choice of Pauli measurements, we get an accurate estimate of $F(\rho,\sigma)$.  

Although we measure only a constant number of Pauli observables, we need to repeat each measurement many times in order to estimate the corresponding expectation value.  The number of repetitions depends on the desired state $\rho$.  In the worst case, it is $O(d)$, but in many cases of practical interest, it is much smaller.  For example, for stabilizer states, the number of repetitions is \textit{constant}, independent of the size of the system; and for the W state, it is only quadratic in the number of qubits $n$.

Even in the worst case, our method requires far fewer resources than full tomography, both in theory and in practice. We demonstrate this by proving lower bounds on the sample complexity of full tomography, and by numerical simulations. 

Finally, we show an analogous method for certifying any unitary quantum channel by estimating the entanglement fidelity.  We discuss applications to benchmarking quantum circuits --- as a special case, our method can certify Clifford circuits in \textit{constant} time, independent of the number of qubits and gates.

\paragraph{Fidelity Estimation.}
The fidelity between our desired pure state $\rho$ and the actual state $\sigma$ is given by~\cite{fidelity}:
\begin{equation}
\begin{split}
F(\rho,\sigma) &= \bigl(\Tr\bigl[(\sqrt{\rho}\sigma\sqrt{\rho})^{1/2}\bigr]\bigr)^2 = \Tr(\rho\sigma).
\end{split}
\end{equation}

We can write $\Tr(\rho\sigma)$ in terms of the Pauli expectation values of $\rho$ and $\sigma$. Let $W_k$ ($k=1, \ldots, d^2$) denote all possible Pauli operators ($n$-fold tensor products of $I$, $\sigma_x$, $\sigma_y$ and $\sigma_z$).  Define the characteristic function $\chi_\rho(k) = \Tr(\rho W_k/\sqrt{d})$, and note that 
\begin{equation}
\label{eqn-tr-rho-sigma}
\Tr(\rho\sigma) = \sum_k \chi_\rho(k) \chi_\sigma(k).
\end{equation}

In general, \eref{eqn-tr-rho-sigma} involves the expectation values of all $d^2$ Pauli operators.  However, it is easy to see that in certain cases fewer Pauli operators are required.  For example, if $\rho$ is a stabilizer state, $\chi_\rho(k)$ takes on values of $\pm 1/\sqrt{d}$ at the $d$ points in the stabilizer group of $\rho$, and vanishes everywhere else. So the sum in \eqref{eqn-tr-rho-sigma} contains only $d$ terms, and one can compute $\Tr(\rho\sigma)$ by measuring only $d$ Pauli operators.  Furthermore, to merely estimate $\Tr(\rho\sigma)$ one only needs to measure a small random subset of these Pauli operators. We will now generalize this strategy to work with an arbitrary pure state $\rho$.

We will construct an estimator for $\Tr(\rho\sigma)$ as follows.  Select $k \in \set{1,\ldots,d^2}$ at random with probability~\cite{prob}
\begin{equation}
\label{eqn-prob-def}
\Pr(k) = (\chi_\rho(k))^2.  
\end{equation}
By measuring the expectation value of the Pauli observable $W_k$, we can estimate $\chi_\sigma(k)$, up to some finite precision which we will discuss later.  For the time being, let us suppose we can measure $\chi_\sigma(k)$ perfectly.  We then construct the estimator 
\begin{equation}
\label{eqn-X-estimator}
X = \chi_\sigma(k) / \chi_\rho(k).  
\end{equation}
It is easy to see that $\EE X = \Tr(\rho\sigma)$ (where $\EE$ denotes the expected value over the random choice of $k$).  

Now say we want to estimate $\Tr(\rho\sigma)$ with some fixed additive error $\varepsilon$ and failure probability $\delta$.  We repeat the above process $\ell = \lceil 1/(\varepsilon^2\delta) \rceil$ times:  we choose $k_1,\ldots,k_\ell$ independently, which give independent estimates $X_1,\ldots,X_\ell$, and we let $Y = \frac{1}{\ell} \sum_{i=1}^\ell X_i$.  By Chebyshev's inequality~\cite{appendix}, $Y$ satisfies
\begin{equation}
\label{eqn-Y}
\Pr[\abs{Y-\Tr(\rho\sigma)} \geq \varepsilon] \leq \delta.
\end{equation}

To complete the description of our method, we show how the ideal ``infinite-precision'' estimator $Y$ can be approximated by an estimator $\tilde{Y}$ that uses a finite number of copies of the state $\sigma$.  Given any choice of $k_1,\ldots,k_\ell$, we proceed as follows.  For each $i=1,\ldots,\ell$, we will use $m_i$ copies of $\sigma$, where we set 
\begin{equation}
\label{eqn-m-choice}
m_i = \biggl\lceil \frac{2}{d\chi_\rho(k_i)^2 \ell\varepsilon^2} \log(2/\delta) \biggr\rceil.
\end{equation}
(Note that $m_i$ depends on $k_i$.)  We measure the Pauli observable $W_{k_i}$ on each of these copies of $\sigma$, and get measurement outcomes $A_{ij} \in \set{1,-1}$ ($j=1,\ldots,m_i$).  Note that $\EE A_{ij} = \sqrt{d}\chi_\sigma(k_i)$ (taking the expectation over the random measurement outcomes).  Let 
\begin{equation}
\tilde{X}_i = \frac{1}{m_i\sqrt{d} \chi_\rho(k_i)} \sum_{j=1}^{m_i} A_{ij}.
\end{equation}
Finally, we let $\tilde{Y} = \frac{1}{\ell} \sum_{i=1}^\ell \tilde{X}_i$.  This is our estimate for $Y$.  (Note that $\EE\tilde{Y} = Y$.)  By Hoeffding's inequality~\cite{appendix}, $\tilde{Y}$ has additive error $\varepsilon$ and failure probability $\delta$:
\begin{equation}
\label{eqn-Ytilde}
\Pr[\abs{\tilde{Y}-Y} \geq \varepsilon] \leq \delta.
\end{equation}

We can then conclude that, with probability $\geq 1-2\delta$, the fidelity $F(\rho,\sigma)$ lies in the range $[\tilde{Y}-2\varepsilon, \tilde{Y}+2\varepsilon]$.  

Our method uses $\ell = \lceil 1/(\varepsilon^2\delta) \rceil$ Pauli observables, independent of the size of the system.  
It requires $m$ copies of the state $\sigma$, where $m = \sum_{i=1}^\ell m_i$. Though this depends on the random choices $k_i$, we have
\begin{equation}
\EE(m_i) = \sum_{k_i} (\chi_\rho(k_i))^2 m_i
 \leq 1 + \frac{2d}{\ell\varepsilon^2} \log(2/\delta),
\end{equation}
and hence the expected number of copies satisfies
\begin{equation}
\label{eqn-m}
\EE(m) \leq 1 + \frac{1}{\varepsilon^2\delta} + \frac{2d}{\varepsilon^2} \log(2/\delta).
\end{equation}
By Markov's inequality, $m$ is unlikely to exceed its expectation by much:  $\Pr(m \geq t\cdot\EE(m)) \leq 1/t$, for all $t\geq 1$.  

\paragraph{Example: the W state.}  Suppose our desired state $\rho$ is the W state, i.e.\ the uniform superposition over computational basis states where a single qubit is $\ket{1}$ and the rest are $\ket{0}$, as previously considered in~\cite{Somma06, Guhne07}. 
To apply our method, we need to sample Pauli operators from the probability distribution \eqref{eqn-prob-def}. 
It is straightforward to give a short formula for these probabilities, and an explicit algorithm that does the sampling in $\poly(n)$ time \cite{appendix}.

The distribution for a W state is quite different from what one would expect for a Haar-random quantum state.  For a random state, one expects most of the Pauli matrices to occur with probability $\sim 1/d^2$; but for the W state, most of the Pauli matrices have probability 0, and all the nonzero probabilities are at least $1/n^2 d$.  This is an example of a \emph{well-conditioned} state.  As we now show, our method requires fewer resources for such states.

\paragraph{Well-conditioned states.}
We say that a state $\rho$ is well-conditioned with parameter $\alpha$ if for all $k$, either $\Tr(\rho W_k) = 0$ or $\abs{\Tr(\rho W_k)} \geq \alpha$. For example, stabilizer states (including the GHZ state) and the W state are well-conditioned with $\alpha=1$ and $\alpha=1/n$, respectively, and Dicke states with $k$ excitations have $\alpha = O(1/n^k)$.
When $\rho$ is well-conditioned, our method requires a smaller number of measurement settings, as well as fewer copies of the actual state $\sigma$.  Note first that the estimator $X$ is bounded: $\abs{X} \leq 1/\alpha$. Now we can use the stronger Hoeffding inequality for \eref{eqn-Y}, and we can choose the number of measurement settings to be $\ell = O\bigl(\frac{\log(1/\delta)}{\alpha^2\varepsilon^2}\bigr)$. Thus, the dependence on $\delta$ is exponentially better, at a cost of a factor of $1/\alpha^2$.

The total number of copies used in the procedure, $m$, is bounded in expectation by \eqref{eqn-m}. For well-conditioned states, we can prove a much stronger bound that holds with certainty: $m_i \leq 1 + \frac{2 \log(2/\delta)}{\alpha^2\ell\varepsilon^2}$, and hence $m \leq O\bigl(\frac{\log(1/\delta)}{\alpha^2\varepsilon^2}\bigr)$.  In particular, when $\rho$ is a stabilizer state, $m$ is \emph{independent} of the size of the system; when $\rho$ is the W state, $m$ is only quadratic in the number of qubits $n$.

\paragraph{Truncating bad events.}
For an arbitrary pure state $\rho$, it is possible to modify our protocol so that $m$ is always bounded by $O\bigl(\frac{1}{\varepsilon^2\delta} + \frac{d \log(1/\delta)}{\varepsilon^2}\bigr)$. The idea is to construct a nearby $\rho'$ which is well-conditioned with $\alpha = O(1/\sqrt{d})$, by truncating small values of $\chi_\rho(k)$. This eliminates the bad choices of $k$ that cause $m$ to be large, at the expense of introducing a small bias into the fidelity estimate \cite{appendix}.  

\paragraph{Dephasing and depolarizing noise.}
Our method also performs better if one makes some mild assumptions about the noise in the system.  For an arbitrary pure state $\rho$, suppose the actual state $\sigma$ is given by $\sigma = \mathcal{E}(\rho)$, where $\mathcal{E}$ is some quantum process that shrinks the characteristic function, i.e., for all $k$, $\abs{\chi_{\mathcal{E}(\rho)}(k)} \leq \abs{\chi_\rho(k)}$.  For example, dephasing and depolarizing noise both do this.  Again, this implies that $\abs{X} \leq 1$, hence we can use a smaller number of measurement settings, $\ell = O\bigl(\frac{\log(1/\delta)}{\varepsilon^2}\bigr)$.

\paragraph{Comparison with full tomography.}
We have shown that it is possible to estimate the fidelity of an arbitrary pure state using Pauli measurements on $O(d)$ copies of the state.  (In this discussion, let us fix the accuracy $\varepsilon$ and failure probability $\delta$ to be constant.)  
How good is this result?  We argue that our protocol is more efficient than full tomography by a factor of $d$. By tomography, we mean any procedure that distinguishes arbitrary quantum states with accuracy $\Delta$, so that for every pair of states $\rho$ and $\sigma$ with $F(\rho,\sigma) \leq 1-\Delta$, the procedure returns different outputs for $\rho$ and $\sigma$.

First, as a toy example, consider what is possible using \textit{arbitrary} quantum operations.  Fidelity estimation of a pure state can then be done with $O(1)$ copies using the swap test~\cite{Buhrman2001}, while full tomography requires $\Omega(d/\poly\log d)$ copies, by Holevo's theorem~\cite{Holevo1973} (see \cite{appendix}).  

In the more realistic situation where only Pauli measurements are allowed (and one cannot perform joint measurements on more than one copy of the state), fidelity estimation uses $O(d)$ copies.  We now prove that full tomography requires at least $\Omega(d^2 / \log d)$ copies.  The idea of the proof is as follows (details in \cite{appendix}).  First, we construct a set of $2^{\Omega(d)}$ quantum states $\ket{\phi_i}$ that are almost orthogonal (for all $i\neq j$, $\abs{\braket{\phi_i}{\phi_j}}^2 < 1- \Delta$), and whose Pauli expectation values are small (for all $i$ and $k$ with $W_k \neq I$, $\abs{\bra{\phi_i}W_k\ket{\phi_i}} \leq \tau\sqrt{\log d}/\sqrt{d}$).  (This is done using repeated applications of Levy's lemma~\cite{appendix,Levy1951}.)  

Now suppose there is some tomography procedure that can distinguish these states, given $t$ copies.  This implies the existence of a classical protocol for transmitting $\Omega(d)$ bits of information over a particular noisy channel $\calE$.  Intuitively, Bob encodes an $\Omega(d)$-bit message $i$ by sending a string of $\pm 1$ bits through the channel $\calE$, in such a way that when Alice receives these bits, they have the same distribution as the measurement outcomes she would have obtained by measuring Pauli observables on the state $\ket{\phi_i}$.  Then Alice uses the tomography procedure to reconstruct $\ket{\phi_i}$ and extract the message $i$.  One can show that the channel $\calE$ has capacity $O\bigl((\log d)/d\bigr)$ (even allowing feedback from Alice to Bob)~\cite{Cover1991}.  Then the converse to Shannon's (classical) noisy coding theorem~\cite{Cover1991} implies that $t \geq \Omega(d^2/\log d)$.

\begin{figure}[t]
\begin{center}
\includegraphics[height=4cm,width=4.25cm]{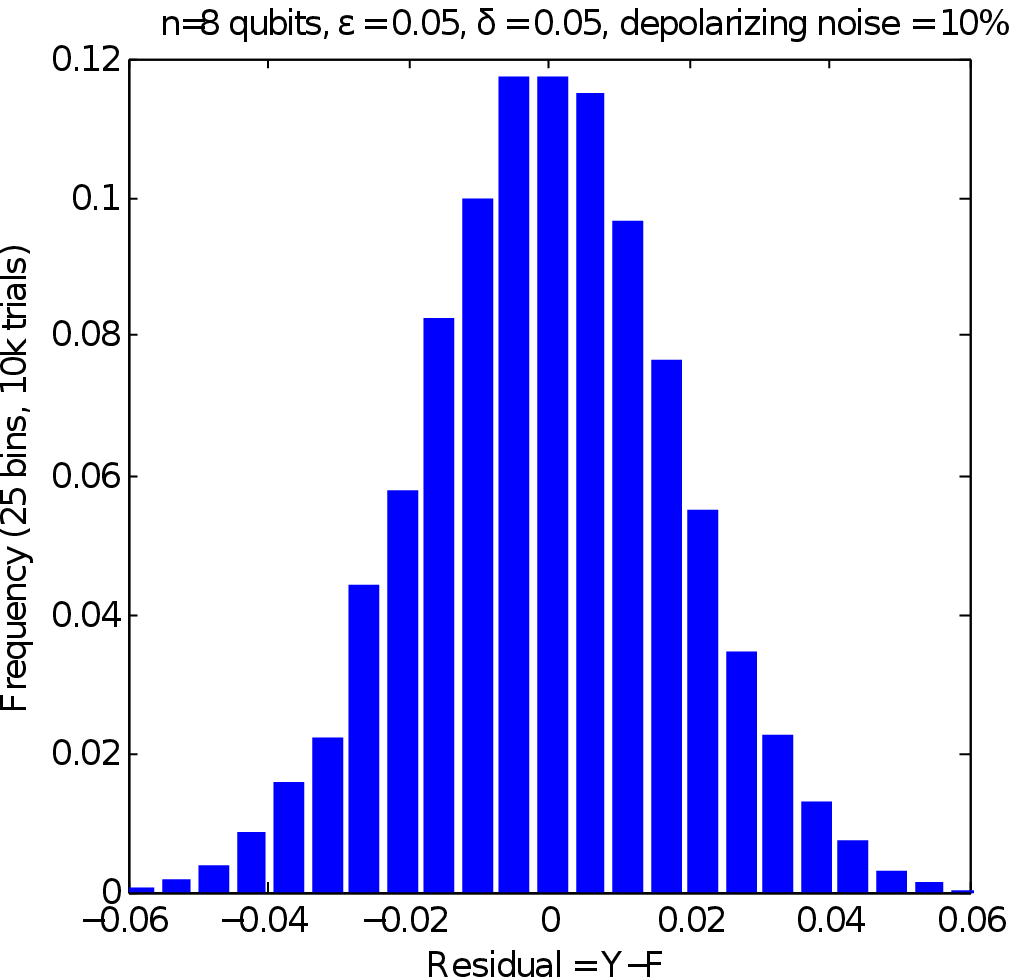}
\includegraphics[height=4cm,width=4.25cm]{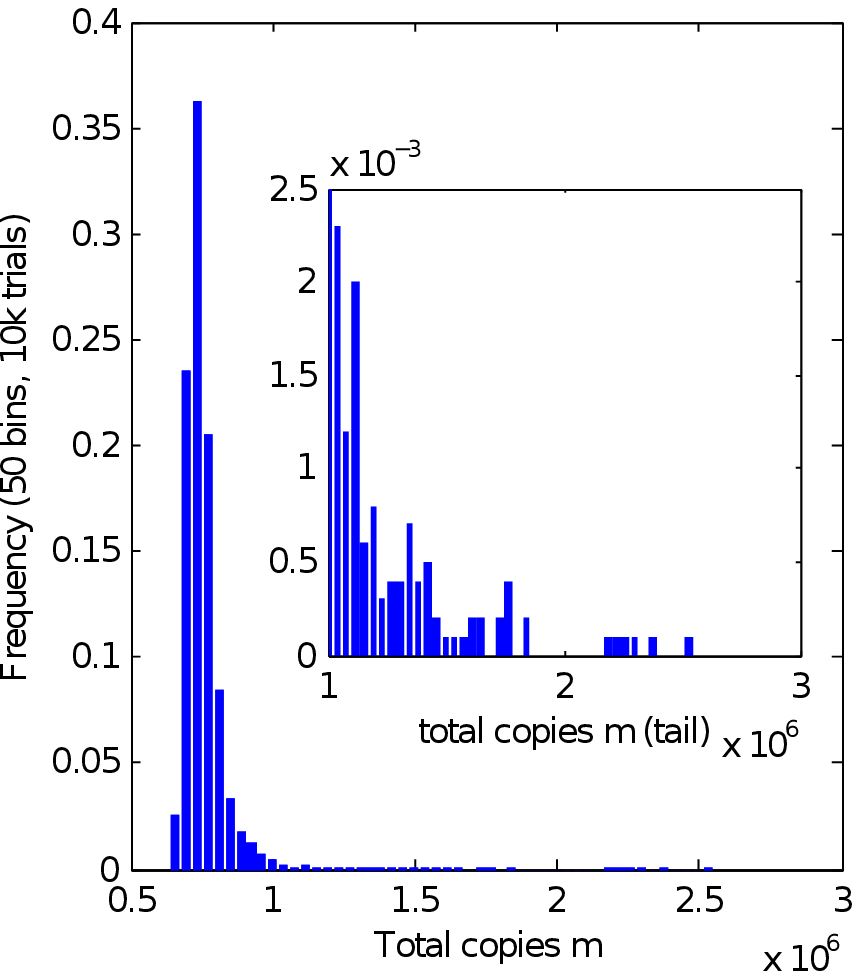}
\caption{\emph{Left:} The residual error has a standard deviation of 1.8\%. \emph{Right:} Most states use only a typical number of copies, with just .1\% of trials using more than four times the expected number of copies, as shown in the inset.
\vspace{-20pt}}
\label{F:fig}
\end{center}
\end{figure}

\paragraph{Numerics.}
In order to evaluate how tight our analysis is for typical states, we simulated our protocol as follows. We sampled Haar-random states of $n=8$ qubits and ran our protocol with $\varepsilon = \delta = .05$ (and $\ell = \frac{1}{\varepsilon^2 \delta}$) where the true state was created by subjecting the ideal state to independent 10\% depolarizing noise. The residual error ($Y-F$) and the total number of copies $m$ are plotted as histograms in Fig.~\ref{F:fig}. We see that the accuracy is always well-behaved, and the total number of copies, excepting a few bad events (for which our truncation procedure applies) is typically close to the average. 

We also compared our method to a recent ion trap experiment, in which an 8-qubit W state was verified using full tomography~\cite{Haffner2005}. Under the plausible assumption that dephasing noise is dominant, we would use our protocol with $\varepsilon = .03$, $\delta = .10$, and $\ell = \lceil\log(1/\delta)/\varepsilon^2\rceil$. Assuming the realistic parameters of $20\,$ms to perform one measurement and $400\,$ms to reconfigure a new measurement basis, we would obtain a fidelity estimate accurate to within $\pm 1.2\%$ using just 80 minutes of experiments and a few seconds of classical processing; this compares very favorably with the 10 \emph{hours} of experiments and one \emph{week} of post-processing carried out in \cite{Haffner2005}.

%
%
%

\paragraph{Extension to channels.} 

We now extend our method to unitary quantum channels.  Let $\calU$ be the desired channel corresponding to some unitary evolution $U$, i.e., $\calU:\: \rho \mapsto U\rho U^\dagger$.  Let $\calE$ be the actual channel.  We will estimate the \emph{entanglement fidelity}, given by $F_{e} = \Tr(\calU^\dagger \calE)/d^2$ (with $\calU$ and $\calE$ treated as matrices acting via left multiplication).

Most of the analysis for channels is exactly analogous to the case of states. The main difference is that we may also \emph{input} a state to the channel as well as choose how to measure at the output. Thus, the characteristic function for a channel $\calE$ is defined by $\chi_\calE(k,k') = \tfrac{1}{d} \Tr\bigl(W_k \calE(W_{k'})\bigr)$,
which depends on two indices. The probability distribution from which we sample indices is analogous: $\Pr(k,k') = \frac{1}{d^2} \bigl[\chi_\calU(k,k')\bigr]^2$, and so is our primary estimator: $X = \chi_\calE(k,k') / \chi_\calU(k,k')$, for which we have $\EE X = F_e$. Now given $\ell$ independent samples from our probability distribution $(k_1,k'_1),\ldots,(k_\ell,k'_\ell)$, we compute $X_1,\ldots,X_\ell$, and let $Y = \frac{1}{\ell} \sum_{i=1}^\ell X_i$. Then choosing $\ell = \lceil 1/(\varepsilon^2\delta) \rceil$ means that $Y$ is an estimate of $F_e$ which is accurate to within $\varepsilon$ with a failure probability at most $\delta$.

The main difference between states and channels comes in how we estimate $X_i$ for a given sample $(k_i,k'_i)$. We will still measure $W_{k_i}$ at the output, but how can we simulate inputing $W_{k'_i}$ into the channel? The key insight is that we can simply sample from states in the eigenbasis of $W_{k'_i}$ and put these states into the channel. Note that these states can always be chosen to be tensor products of local Pauli eigenstates, so no entangling gates are required. 

The total number of uses of the channel is bounded in expectation by $\EE(m) = O\bigl(\frac{1}{\varepsilon^2\delta} + \frac{d^2}{\varepsilon^2} \log(1/\delta)\bigr)$. Statements about well-conditioned channels and truncation also hold in analogy with states~\cite{appendix}.

\paragraph{Benchmarking quantum circuits.}
One application of the above protocol is to evaluate experimental implementations of large quantum circuits: our method allows one to directly measure the entanglement fidelity and average fidelity of the entire circuit, rather than inferring it from tomography performed on individual gates.  This is important because as circuits scale up, correlated noise potentially becomes an issue (c.f.\ Ref.~\cite{Monz2011}).

The relationship between $F_e$ and the Haar-average fidelity is captured by the formula~\cite{Horodecki1999a}
\begin{align}
	F_{\mathrm{avg}} = \int \mathrm{d}\psi F\bigl(\calU(\psi),\calE(\psi)\bigr) = \tfrac{d}{d+1} F_e + \tfrac{1}{d+1}\,.
\end{align}
Thus, our method also gives us a direct measure of the typical performance of the channel, similar to what is achieved in other random benchmarking schemes~\cite{Knill2008, Magesan2010, Emerson2007}. Moreover, one can also prove that the worst-case behavior (as quantified by the diamond norm~\cite{Aharonov1998}) is bounded by $4 d \sqrt{1-F_e}$~\cite{Beigi2011}, so that for small high-fidelity gates, average and worst-case behavior nearly coincide.

\paragraph{Clifford circuits.}
Clifford circuits (those consisting of controlled-NOT, Hadamard and phase gates) are key components in many schemes for quantum error-correction, and become universal for quantum computation when augmented with certain state preparations~\cite{Gottesman1999, Bravyi2005}. 
For a Clifford circuit $\calU$, the characteristic function is given by $\chi_\calU(k,k') = 1$ (when $W_k = \calU(W_{k'})$) and 0 otherwise.  Sampling only requires that we pick $k' \in \set{1,\ldots,d^2}$ uniformly at random, then use the Gottesman-Knill theorem to efficiently compute the $k$ such that $W_k = \calU(W_{k'})$.  Clifford circuits are well-conditioned, so our method needs fewer measurement settings and uses of the channel, namely $\ell \leq O\bigl(\frac{1}{\varepsilon^2} \log(1/\delta)\bigr)$ and $m \leq O\bigl(\frac{1}{\varepsilon^2} \log(1/\delta)\bigr)$, which is \emph{independent} of the number of qubits and gates (see also Ref.~\cite{Low2009a}).

\paragraph{Outlook.}

We have presented a general method for certifying pure states and unitary quantum channels, which requires only Pauli measurements and is faster than full tomography by a factor of $d$.  In common cases such as stabilizer states, the W state, and Clifford circuits, our method requires even fewer resources (constant or polynomial in the number of qubits), and it provides an easy recipe to generalize beyond these examples.  

Looking beyond fidelity estimation, it would be interesting to directly estimate and bound an entanglement measure \cite{blume-kohout10}, which would obviate the need for an entanglement witness. One may also compare our method with recent proposals for tomography for restricted classes of quantum states \cite{Gross2010, Shabani2011, mps-tomog, permut-tomog}.  
Another important direction is to find better techniques for sampling the importance-weighting distribution $\Pr(k)$ for different classes of states.

\acknowledgements
We thank D.~Gross, J.~Preskill and T.~Monz for helpful discussions. YKL was supported by NIST Grant No. 60NANB10D262, and STF by NSF Grant No. PHY-0803371 and ARO Grant No. W911NF-09-1-0442. 

We would also like to note that M. da Silva, O. Landon-Cardinal and D. Poulin~\cite{da-Silva2011} have recently and independently obtained similar results.

\vspace{-10pt}
\bibliography{cert,refs,books}

\begin{thebibliography}{30}
\bibitem [{pro()}]{prob}%
  \BibitemOpen
  \href@noop {} {}\bibinfo {note} {These probabilities are normalized since
  $\Tr(\rho^2) = 1$.}\BibitemShut {Stop}%
\bibitem [{app()}]{appendix}%
  \BibitemOpen
  \href@noop {} {}\bibinfo {note} {We defer some details to the
  appendices.}\BibitemShut {Stop}%
\end{thebibliography}

\clearpage

\appendix
\section{Bounding the Failure Probabilities}

To show \eref{eqn-Y}, observe that the variance of each individual estimator $X_i$ is not too large,
\begin{align}
	\Var(X_i) & = \EE(X_i^2) - (\EE X_i)^2 \\
	& = \sum_k [\chi_\sigma(k)]^2-[\Tr(\rho\sigma)]^2 \\
	& = \Tr(\sigma^2) - [\Tr(\rho\sigma)]^2 \leq 1.
\end{align}
This implies that $\Var(Y) \leq 1/\ell$. Hence, by Chebyshev's inequality, 
\begin{equation}
\label{eqn-chebyshev}
\Pr\Bigl[\abs{Y - \Tr(\rho\sigma)} \geq \lambda/\sqrt{\ell}\Bigr] \leq \frac{1}{\lambda^2}.
\end{equation}
Then set $\lambda = 1/\sqrt{\delta}$ and $\ell = \lceil 1/(\varepsilon^2\delta) \rceil$.

To show Eq.~\eqref{eqn-Ytilde}, we use Hoeffding's inequality, which says that for all $\varepsilon >0$,
\begin{equation}
	\Pr\bigl[\abs{\tilde{Y}-Y} \geq \varepsilon\bigr] \leq 2\exp(-2\varepsilon^2/C),
\end{equation}
where 
\begin{equation}
C = \sum_{i=1}^\ell \sum_{j=1}^{m_i} (2c_i)^2, \quad
c_i = \frac{1}{\ell m_i\sqrt{d} \chi_\rho(k_i)}.
\end{equation}
Setting $m_i$ as in \eref{eqn-m-choice}, we get 
\begin{equation}
C = \sum_{i=1}^\ell \frac{4}{\ell^2 m_i d\chi_\rho(k_i)^2}
 \leq \frac{2\varepsilon^2}{\log(2/\delta)},
\end{equation}
hence the failure probability is $\leq \delta$, as claimed.

\section{Efficient sampling for the W state}

Recall the definition of the W state,
\begin{align}
	\ket W = \frac{1}{\sqrt{n}}\sum_{\abs{\vec{b}}=1} \ket{\vec{b}}\,,
\end{align}
where the sum is over all $n$-bit strings $\vec{b}$ with Hamming weight $\abs{\vec{b}}=1$. We can factor any $n$-qubit Pauli operator into a tensor product of local Pauli $\sigma_x$ operators times a tensor product of local Pauli $\sigma_z$ operators (up to an irrelevant phase). Our probability distribution follows from the definition in \eref{eqn-prob-def},
\begin{align}
	p(\vec{j},\vec{k}) = \Pr(\sigma_x^{\vec{j}} \sigma_z^{\vec{k}}) = \frac{1}{d} \abs{\bra W \sigma_x^{\vec{j}} \sigma_z^{\vec{k}} \ket W}^2\,,
\end{align}
where we denote the tensor product by a bit string in the exponent. (Thus, for example, $\sigma_x^{110}\sigma_z^{011} = \sigma_x \otimes \sigma_y \otimes \sigma_z$, up to an irrelevant phase.)
\begin{align}
	p(\vec{j},\vec{k}) 
	& = \frac{1}{n^2 d} \biggl\lvert\sum_{\abs{\vec{a}}=\abs{\vec{b}}=1}
		\bra{\vec{a}} \sigma_x^{\vec{j}} \sigma_z^{\vec{k}} \ket{\vec{b}}\biggr\rvert^2 \\
	& = \frac{1}{n^2 d} \biggl\lvert \sum_{\abs{\vec{a}}=\abs{\vec{b}}=1} 
		(-1)^{\vec{b}\cdot \vec{k}}\delta_{\vec{a},\vec{b}+\vec{j}} \biggr\rvert^2\,,
\end{align}
where the arithmetic in the delta function is modulo 2. The delta function tells us that the tensor product over $\sigma_x$ must only contain either $0$ or $2$ factors of $\sigma_x$ only; all other terms have zero probability. Let's separate out the case where there are no $\sigma_x$ operators from when there are two.  If there are none, then
\begin{align}
	p(\vec{0},\vec{k}) 
	& = \frac{1}{n^2 d} \biggl\lvert \sum_{\abs{\vec{b}}=1} (-1)^{\vec{b}\cdot \vec{k}} \biggr\rvert^2 = \frac{1}{n^2 d} \biggl\lvert \sum_{i=1}^n (-1)^{k_i} \biggr\rvert^2 \\
	& = \frac{1}{n^2 d} \bigl( n-2\abs{\vec{k}}\bigr)^2\,.
\end{align}
If $\vec{j}$ has weight 2, then the summand reduces to only two terms, since flipping two bits in the weight-1 string $\vec{b}$ will (with two exceptions) increase the weight, making it orthogonal to the weight-1 string $\vec{a}$.
\begin{align}
	p(\vec{j},\vec{k}) & = \frac{1}{n^2 d} \biggl\lvert \sum_{\abs{\vec{a}}=\abs{\vec{b}}=1} (-1)^{\vec{b}\cdot \vec{k}}\delta_{\vec{a},\vec{b}+\vec{j}} \biggr\rvert^2\\
	& =\frac{1}{n^2 d} \bigl( 1+(-1)^{\vec{j}\cdot\vec{k}} \bigr)^2\,.
\end{align}
This is clearly either $0$ or $4/n^2d$ depending on $\vec{j}\cdot\vec{k} \bmod 2$. To summarize, we have the following formula for the probabilities
\begin{align}
	p(\vec{j},\vec{k}) = \begin{cases}\frac{1}{n^2 d} \bigl( n-2\abs{\vec{k}}\bigr)^2 & \mbox{ if } \vec{j} = \vec{0} \\
\frac{4}{n^2 d} & \mbox{ if } \abs{\vec{j}} = 2 \, , \, \vec{j}\cdot\vec{k} = 0\\
0 & \mbox{ otherwise, } \end{cases}
\end{align}
where again, the dot product $\vec{j}\cdot\vec{k}$ is taken mod 2.

Given this formula, we have the following simple procedure to sample from this distribution. The procedure consists of two steps. First, flip a weighted coin to see if you are in the first or the second branch. The total weight in the first branch (with $\vec{j}=0$) is $1/n$, a fact that follows from some simple binomial identities, or by directly computing the weight in the second branch. If we are in this first branch, then all strings $\vec{k}$ of a given Hamming weight are equally probable. We can sample from this by first picking the weight $w = \abs{\vec{k}}$ from the normalized distribution
\begin{align}
	q(w) = \frac{1}{n d} {n \choose w} \bigl( n-2w\bigr)^2 \,.
\end{align}
Since this distribution only has $n$ outcomes, we can sample from it efficiently in $n$. Then we just choose a random bit string with the given sampled weight.  Now consider that we are in the second branch after the initial coin flip. Then we choose uniformly from all ${n \choose 2}$ bit strings of length $n$ containing exactly two ones, and this defines $\vec{j}$. Then we pick a bit string $\vec{k}$ by choosing a uniformly random bit string of length $n-1$, and we take (say) the first bit and copy it between the two sites in $\vec{k}$ which are supported by $\vec{j}$ to enforce the condition $\vec{j}\cdot\vec{k}=0 \bmod 2$ and distribute the remaining random bits over the rest of $\vec{k}$ sequentially. 

\section{Truncating Bad Events}

The modified procedure is as follows:  construct a new state $\rho_1$ by defining its characteristic function to be 
\begin{equation}
\chi_{\rho_1}(k) = \begin{cases}
  \chi_\rho(k) & \text{ if $\abs{\chi_\rho(k)} \geq \beta/d$,} \\
  0 & \text{ otherwise.}
\end{cases}
\end{equation}
Define $\rho_2 = \rho_1 / \norm{\rho_1}_2$, where $\norm{\rho}_2 = \sqrt{\Tr(\rho^2)}$ is the Schatten 2-norm.  Then perform our original certification procedure using $\rho_2$, to estimate $\Tr(\rho_2\sigma)$.  Actually, note that $\rho_2$ may not be a density matrix (it may not be positive semidefinite with trace 1); nonetheless, it satisfies $\Tr(\rho_2^2) = 1$, so the certification procedure makes sense.  

We can bound $m$ as follows:  note that, for all $k$, either $\chi_{\rho_2}(k) = 0$, or $\abs{\chi_{\rho_2}(k)} \geq \abs{\chi_{\rho_1}(k)} \geq \beta/d$.  Then, with probability 1, we have 
$m_i \leq 1 + \frac{2d}{\beta^2\ell\varepsilon^2} \log(2/\delta)$ 
and 
$m \leq 1 + \frac{1}{\varepsilon^2\delta} + \frac{2d}{\beta^2\varepsilon^2} \log(2/\delta)$.

We claim that $\Tr(\rho_2\sigma)$ gives us an estimate of $\Tr(\rho\sigma)$, with some bias that is not too large.  Clearly, 
\begin{equation}
\abs{\Tr(\rho_2\sigma) - \Tr(\rho\sigma)} \leq \norm{\rho_2-\rho}_2, 
\end{equation}
and the quantity on the right-hand side can be calculated explicitly, given knowledge of $\rho$.  In the worst case, we claim that $\norm{\rho_2-\rho}_2 \leq 2\beta$.  To see this, note that $\norm{\rho_1-\rho}_2 \leq \beta$, and $1-\beta \leq \norm{\rho_1}_2 \leq 1$, hence $\norm{\rho_2-\rho_1}_2 \leq \beta$.

\section{Lower Bound for Tomography}

As a toy example, consider the situation where we can perform arbitrary quantum operations.  In this setting, full tomography of a pure state with constant accuracy requires at least $\Omega(d)$ copies (up to log factors); this follows from the existence of sets of $2^{\Omega(d)}$ almost-orthogonal pure states~\cite{Buhrman2001}, and Holevo's theorem~\cite{Holevo1973}.  Note that this lower bound is tight:  full tomography can be done with $O(d)$ copies, by using random POVM measurements \cite[][Thm.~3]{Sen2006} to perform state discrimination on an $\varepsilon$-net of pure states \cite[][Lemma II.4]{Hayden2004}.  

Now consider the more realistic situation where only Pauli measurements are allowed.  We prove that full tomography using Pauli measurements requires $\Omega(d^2/\log d)$ copies of the state.

\textit{First step.}  We want to construct a large set of nearly-orthogonal quantum states that have small Pauli expectation values.  To do this, we will use the following lemma:  
\begin{lemma}
Fix any states $\ket{\phi_1},\ldots,\ket{\phi_s} \in \CC^d$, where $s \leq 2^{cd}$ and $c$ is some constant.  Then there exists a state $\ket{\psi} \in \CC^d$ such that:
\begin{equation}
\label{eqn-orthog}
\forall\, i \in \set{1,\ldots,s}, \ \abs{\braket{\phi_i}{\psi}} \leq \varepsilon, 
\end{equation}
\begin{equation}
\label{eqn-pauli-small}
\forall\, W_k \neq I, \ \abs{\bra{\psi}W_k\ket{\psi}} \leq \tau\sqrt{\log d}/\sqrt{d}.
\end{equation}
Here $\varepsilon = \sqrt{9\pi^3(\log 2)c}$ and $\tau = \sqrt{72\pi^3}$.
\end{lemma}

The proof of the lemma is as follows.  Choose $\psi$ to be a Haar-random vector in $S^{d-1}$.  We claim that (\ref{eqn-orthog}) and (\ref{eqn-pauli-small}) are satisfied with high probability.  

First, for each $i$, observe that $\braket{\phi_i}{\psi}$ is a smooth function of $\psi$, with Lipschitz coefficient $\eta = 1$:
\begin{equation}
\bigl\lvert \braket{\phi_i}{\psi} - \braket{\phi_i}{\psi'} \bigr\rvert \leq \norm{\psi-\psi'}_2.
\end{equation}
By symmetry, $\EE \braket{\phi_i}{\psi} = 0$.  So by Levy's lemma~\cite{Levy1951}, 
\begin{equation}
\Pr\bigl[\abs{\braket{\phi_i}{\psi}} \geq \varepsilon\bigr] \leq 4\exp(-C_1 d\varepsilon^2/\eta^2), 
\end{equation}
where $C_1 = 2/9\pi^3$.  Taking the union bound over all $i$, we get that 
\begin{equation}
\begin{split}
\Pr&[\text{\eref{eqn-orthog} fails for some $i$}] \\
 &\quad \leq 4\exp(cd(\log 2) - C_1 d\varepsilon^2) \\
 &\quad = 4\exp(-cd(\log 2)) = 4\cdot 2^{-cd}.
\end{split}
\end{equation}

Next, for each $k$, observe that $\bra{\psi}W_k\ket{\psi}$ is a smooth function of $\psi$, with Lipschitz coefficient $\eta = 2$:
\begin{equation}
\begin{split}
\bigl\lvert \bra{\psi}&W_k\ket{\psi} - \bra{\psi'}W_k\ket{\psi'} \bigr\rvert \\
 &\leq \bigl\lvert \bra{\psi}W_k \bigl[ \ket{\psi}-\ket{\psi'} \bigr] \bigr\rvert
     + \bigl\lvert \bigl[ \bra{\psi}-\bra{\psi'} \bigr] W_k \ket{\psi'} \bigr\rvert \\
 &\leq 2\norm{\psi-\psi'}_2.
\end{split}
\end{equation}
By symmetry, $\EE \bra{\psi}W_k\ket{\psi} = 0$.  So by Levy's lemma~\cite{Levy1951}, 
\begin{equation}
\begin{split}
	\Pr\Bigl[\abs{\bra{\psi}W_k\ket{\psi}} \geq \tau&\sqrt{\log d}/\sqrt{d}\Bigr] \leq \\
	&  4\exp(-C_1 \tau^2(\log d)/\eta^2), 
\end{split}
\end{equation}
where $C_1 = 2/9\pi^3$.  Taking the union bound over all $k$, we get that 
\begin{equation}
\begin{split}
\Pr&\bigl[\text{\eref{eqn-pauli-small} fails for some $k$}\bigr] \\
 &\quad \leq 4\exp(2\log d - C_1 \tau^2(\log d)/4) \\
 &\quad = 4\exp(-2\log d) = 4/d^2.
\end{split}
\end{equation}
This proves the lemma.

By applying the above lemma repeatedly, we can construct a set of $2^{\Omega(d)}$ quantum states $\ket{\phi_i}$ that are almost orthogonal (for all $i\neq j$, $\abs{\braket{\phi_i}{\phi_j}}^2 < 1 - \Delta$), and whose Pauli expectation values are small (for all $i$ and $k$ with $W_k \neq I$, $\abs{\bra{\phi_i}W_k\ket{\phi_i}} \leq \tau\sqrt{\log d}/\sqrt{d}$).  

\textit{Second step.}  Suppose there is some tomography procedure that can distinguish among the states $\ket{\phi_i}$, given $t$ copies.  We now construct a classical protocol for transmitting $\Omega(d)$ bits of information over a particular noisy channel $\calE$.  

Let $\mathcal{E}$ be the classical channel that takes a bit $b\in\set{1,-1}$ and outputs a bit $b'\in\set{1,-1}$, where with probability $\tau\sqrt{\log d}/\sqrt{d}$, the channel sets $b' = b$, and with probability $1-\tau\sqrt{\log d}/\sqrt{d}$, the channel chooses $b'\in\set{1,-1}$ uniformly at random.  Using the tomography procedure, we will show how to send messages over this channel (together with a noiseless feedback channel).

Say Bob wants to send $O(d)$ bits to Alice.  He associates the message with a state $\ket{\phi_i}$.  Alice runs the tomography procedure. When she wants to measure some Pauli matrix $W_k$, she sends $k$ to Bob (over the noiseless feedback channel).  Bob chooses a random $b\in\set{1,-1}$ with expectation value $\bra{\phi_i}W_k\ket{\phi_i} \cdot\sqrt{d}/\tau\sqrt{\log d}$, and sends $b$ through the channel $\mathcal{E}$ to Alice.  Alice receives $b'$, which has expectation value $\bra{\phi_i}W_k\ket{\phi_i}$.

For tomography using $t$ copies, Bob sends $t$ bits through the channel $\mathcal{E}$ (in addition to the feedback bits sent by Alice).  But $\mathcal{E}$ is simply the binary symmetric channel, which has capacity $\leq \tau^2(\log d)/d$.  Furthermore, feedback does not increase its capacity~\cite{Cover1991}.  So, by the converse to Shannon's (classical) noisy coding theorem~\cite{Cover1991}, Bob must use the channel at least $\Omega(d^2/\log d)$ times to send $\Omega(d)$ bits.  Hence $t \geq \Omega(d^2/\log d)$.

\section{Estimating entanglement fidelity for channels}

Here we give a detailed description of our method for certifying quantum channels.  Let $\CC^{d\times d}_H$ denote the set of Hermitian matrices in $\CC^{d\times d}$.
We will view $\CC^{d\times d}_H$ as a vector space, with Hilbert-Schmidt inner product $\Tr(A^\dagger B)$.  We use round bra-kets to denote this:  $\Ket{A}$ is a vector, $\Bra{B}$ is an adjoint vector, and $\Braket{A}{B} = \Tr(A^\dagger B)$ is an inner product.  

Let $\calL(\CC^{d\times d}_H, \CC^{d\times d}_H)$ be the vector space of all linear maps from $\CC^{d\times d}_H$ to $\CC^{d\times d}_H$, again with Hilbert-Schmidt inner product $\Tr(\mathcal{A}^\dagger \mathcal{B})$.  Now recall the Pauli matrices $\Ket{W_k} \in \CC^{d\times d}_H$ ($k=1,\ldots,d^2$).  Note that $\tfrac{1}{d} \Ket{W_k} \Bra{W_{k'}}$ ($k,k'\in\set{1,\ldots,d^2}$) form an orthonormal basis for $\calL(\CC^{d\times d}_H, \CC^{d\times d}_H)$.  For any channel $\calE \in \calL(\CC^{d\times d}_H, \CC^{d\times d}_H)$, we define its characteristic function to be 
\begin{equation}
\begin{split}
\chi_\calE(k,k')
 &= \Tr\Bigl[ \Bigl[\tfrac{1}{d} \Ket{W_k} \Bra{W_{k'}}\Bigr]^\dagger \calE \Bigr] \\
 &= \tfrac{1}{d} \Bra{W_k} \calE \Ket{W_{k'}} = \tfrac{1}{d} \Tr(W_k^\dagger \calE(W_{k'})).
\end{split}
\end{equation}
(Note that $\chi_\calE(k,k')$ is real, since $W_k$ and $\calE(W_{k'})$ are Hermitian.)  Then 
\begin{equation}
\calE = \tfrac{1}{d} \sum_{k,k'} \chi_\calE(k,k') \Ket{W_k} \Bra{W_{k'}},
\end{equation}
and the overlap between $\calU$ and $\calE$ is given by 
\begin{equation}
\Tr(\calU^\dagger \calE) = \sum_{k,k'} \chi_\calU(k,k') \chi_\calE(k,k').
\end{equation}
Note that for any channel $\calE$, $0 \leq \Tr(\calE^\dagger \calE) \leq d^2$, and since $\calU$ is a unitary channel, $\Tr(\calU^\dagger \calU) = d^2$.  This implies $\abs{\Tr(\calU^\dagger \calE)} \leq d^2$.  We will be interested in estimating $\Tr(\calU^\dagger \calE)/d^2$ up to an additive error of size $\varepsilon$.

We will construct an estimator for $\Tr(\calU^\dagger \calE)/d^2$ as follows.  Select $(k,k') \in \set{1,\ldots,d^2}^2$ at random with probability 
\begin{equation}
\label{eqn-channel-weights}
\Pr(k,k') = \frac{1}{d^2} \bigl[\chi_\calU(k,k')\bigr]^2.  
\end{equation}
(Note that these probabilities are normalized, since $\Tr(\calU^\dagger \calU) = d^2$.)  We can estimate $\chi_\calE(k,k')$, up to some finite precision, by preparing eigenstates of $W_{k'}$, applying the channel $\calE$, and then measuring the observable $W_k$; we will discuss this below.  We then compute the quantity 
\begin{equation}
X = \chi_\calE(k,k') / \chi_\calU(k,k').  
\end{equation}
It is easy to see that $\EE X = \Tr(\calU^\dagger \calE)/d^2$.  

We want an estimate with additive error $\varepsilon$ and failure probability $\delta$, so we repeat the above process $\ell = \lceil 1/(\varepsilon^2\delta) \rceil$ times:  we choose $(k_1,k'_1),\ldots,(k_\ell,k'_\ell)$ independently, which give independent estimates $X_1,\ldots,X_\ell$, and we let $Y = \frac{1}{\ell} \sum_{i=1}^\ell X_i$.  
(Note that the number of Pauli observables $\ell$ is independent of the size of the system.)  By Chebyshev's inequality,
\begin{equation}
\label{eqn-Y-2}
\Pr\bigl[\abs{Y - \Tr(\calU^\dagger\calE)/d^2} \geq \varepsilon\bigl] \leq \delta.
\end{equation}

Finally, we describe how to estimate $Y$ from a finite number of uses of the channel $\calE$.  Fix any choice of $(k_i,k'_i)$ for $i=1,\ldots,\ell$.  We then estimate $Y$ as follows.  For each $i=1,\ldots,\ell$:
\begin{itemize}
  \item Choose some eigenbasis for the Pauli matrix $W_{k'_i}$, call it $\ket{\phi^i_a}$ ($a=1,\ldots,d$), and let $\lambda^i_a \in \set{1,-1}$ be the corresponding eigenvalues.  (Note that one can choose the $\ket{\phi^i_a}$ to be tensor products of single-qubit Pauli eigenstates.)  
  \item Let 
\begin{equation}
m_i = \biggl\lceil \frac{4}{\chi_\calU(k_i,k'_i)^2 \ell\varepsilon^2} \log(4/\delta) \biggr\rceil.
\end{equation}
  \item For each $j=1,\ldots,m_i$:  choose some $a_{ij}\in\set{1,\ldots,d}$ uniformly at random, prepare the state $\ket{\phi^i_{a_{ij}}}$, apply the channel $\calE$, and measure the Pauli observable $W_{k_i}$, to get an outcome $A_{ij}\in\set{1,-1}$; finally, let $B_{ij} = \lambda^i_{a_{ij}} A_{ij}$.
\end{itemize}
Note that 
\begin{equation}
\begin{split}
\EE B_{ij} &= \tfrac{1}{d} \sum_{a_{ij}=1}^d \lambda^i_{a_{ij}} \Tr(W_{k_i} \calE(\ket{\phi^i_{a_{ij}}}\bra{\phi^i_{a_{ij}}})) \\
 &= \tfrac{1}{d} \Tr(W_{k_i} \calE(W_{k'_i})) = \chi_\calE(k_i,k'_i).
\end{split}
\end{equation}
Let 
\begin{equation}
\tilde{X}_i = \frac{1}{\chi_\calU(k_i,k'_i) m_i} \sum_{j=1}^{m_i} B_{ij}.
\end{equation}
Finally, we let $\tilde{Y} = \frac{1}{\ell} \sum_{i=1}^\ell \tilde{X}_i$.  This is our estimate for $Y$; note that $\EE\tilde{Y} = Y$.  By Hoeffding's inequality, 
\begin{equation}
\label{eqn-Ytilde-2}
\Pr\bigl[\abs{\tilde{Y}-Y} \geq \varepsilon\bigr] \leq \delta.
\end{equation}

This procedure uses the channel $\calE$ a total of $m$ times, where $m = \sum_{i=1}^\ell m_i$.  This number depends on the random choice of $(k_i,k'_i)$ ($i=1,\ldots,\ell$).  We can bound it in expectation:  we have 
\begin{equation}
\EE(m_i) = \sum_{k_i,k'_i} \tfrac{1}{d^2} \chi_\calU(k_i,k'_i)^2 m_i
 \leq 1 + \frac{4d^2}{\ell\varepsilon^2} \log(4/\delta),
\end{equation}
hence
\begin{equation}
\label{eqn-m-2}
\EE(m) \leq 1 + \frac{1}{\varepsilon^2\delta} + \frac{4d^2}{\varepsilon^2} \log(4/\delta).
\end{equation}
Then use Markov's inequality:  $\Pr(m \geq t\cdot\EE(m)) \leq 1/t$, for all $t\geq 1$.  

It remains to prove \eqref{eqn-Y-2} and \eqref{eqn-Ytilde-2}, bounding the failure probability.  To show \eqref{eqn-Y-2}, note that the variance of each $X_i$ is not too large:
\begin{equation}
\begin{split}
\Var(X_i) &= \EE(X_i^2) - (\EE X_i)^2 \\
 &= \sum_{k,k'} \frac{1}{d^2} \chi_\calE(k,k')^2 - \frac{1}{d^4} \Tr(\calU^\dagger\calE)^2 \\
 &= \frac{1}{d^2} \Tr(\calE^\dagger \calE) - \frac{1}{d^4} \Tr(\calU^\dagger\calE)^2 \leq 1.
\end{split}
\end{equation}
Then $\Var(Y) \leq 1/\ell$, so by Chebyshev's inequality, 
\begin{equation}
\label{eqn-chebyshev-2}
\Pr\bigl[\abs{Y - (1/d^2) \Tr(\calU^\dagger\calE)} \geq \tfrac{\lambda}{\sqrt{\ell}}\bigr] \leq \tfrac{1}{\lambda^2}.
\end{equation}
Now set $\lambda = 1/\sqrt{\delta}$ and $\ell = \lceil 1/(\varepsilon^2\delta) \rceil$.

To show \eqref{eqn-Ytilde-2}, we use Hoeffding's inequality:  for any $\varepsilon>0$, 
\begin{equation}
\Pr\bigl[\abs{\tilde{Y}-Y} \geq \varepsilon\bigr] \leq 2\exp(-2\varepsilon^2/C),
\end{equation}
where 
\begin{equation}
C = \sum_{i=1}^\ell \sum_{j=1}^{m_i} (2c_i)^2, \quad
c_i = \frac{1}{\ell \chi_\calU(k_i,k'_i) m_i}.
\end{equation}
We have 
\begin{equation}
C = \sum_{i=1}^\ell \frac{4}{\ell^2 \chi_\calU(k_i,k'_i)^2 m_i}
 \leq \frac{\varepsilon^2}{\log(4/\delta)},
\end{equation}
hence the failure probability is $\leq \delta$ as desired.

\end{document}